\documentclass[journal]{IEEEtran}

\usepackage{amsmath,amsfonts}
\usepackage{algorithmic}
\usepackage{algorithm}
\usepackage{array}
\usepackage[caption=false,font=normalsize,labelfont=sf,textfont=sf]{subfig}
\usepackage{textcomp}
\usepackage{stfloats}
\usepackage{url}
\usepackage{verbatim}
\usepackage{graphicx}
\usepackage{multirow}
\usepackage{cite}
\usepackage{makecell}
\usepackage{comment}

\newcommand{\bs}[1]{\boldsymbol{#1}}
\newcommand{\mci}[1]{\multicolumn{1}{c}{#1}}
\newcommand{\mcii}[1]{\multicolumn{2}{c}{#1}}
\newcommand{\uj}{\mathrm{j}}

\begin{document}

\title{Modal Energy for Power System Analysis: Definitions and Requirements}

\author{Jiahao Liu,~\IEEEmembership{IEEE~Member}, and Federico Milano,~\IEEEmembership{IEEE~Fellow}        
        % Students with Similar Topics?,~\IEEEmembership{IEEE~Student~Member}
        % <-this % stops a space
\thanks{J.~Liu and F.~Milano are with the School of Electrical and Electronic
Engineering, University College Dublin, Belfield Campus, D04V1W8, Ireland.
E-mails: jiahao.liu@ucd.ie, federico.milano@ucd.ie}
\thanks{This work is supported by the Sustainable Energy Authority of Ireland
(SEAI) by funding J.~Liu and F.~Milano under the project FRESLIPS,
Grant No. RDD/00681.}
% This paper was produced by the IEEE Publication Technology Group.  They are in Piscataway, NJ.}% <-this % stops a space
% \thanks{Manuscript received April 19, 2021; revised August 16, 2021.}
\vspace{-9mm}
}

\maketitle

\begin{abstract}
Modal energy provides information complementary to and based on conventional eigenvalues and participation factors for power system modal analysis.  However, modal energy definition is not unique.  This letter clarifies the definitions and applicability of mainstream modal energy approaches, focusing on their mappings to eigenvalues and to the total system energy.  It is shown that these mappings hold only under restrictive conditions, notably system normality, which limits their applicability in inverter-dominated power systems.
\end{abstract}

\vspace{-3mm}

% \section*{Notations on Operator}
% Let $\chi$ is a variable.

% \begin{table}[h]
% \centering
% \normalsize
% \begin{tabular}{p{0.2cm} p{3.2cm} p{0.2cm} p{3.2cm}}
%     $\bs{\chi}$ & Vector or matrix & $\bar{\chi}$ & Complex value \\
%     $\dot{\chi}$ & time derivative & $\chi^{*}$ & Complex conjugate  \\
%     $\chi^{T}$ & Transpose & $\chi^{H}$ & Hermitian transpose  \\
% \end{tabular}
% \end{table}

\vspace{-5mm}

\section{Introduction}

\IEEEPARstart{P}{ower} system small-signal stability is commonly assessed using modal analysis, with eigenvalues and participation factors as indices \cite{Gibbard2015, Milano2021}.  Modal energy has been proposed as a tool to evaluate the amount of energy associated with each mode and how this energy is dissipated.
However, unlike eigenvalues, modal energy does not have a unique definition.  In \cite{MacFarlane1969}, system power is decomposed into modal power, and the relationship between modal power and modal energy is later established through eigenvectors in \cite{Hamdan1986}.  More recent definitions are based on participation factors (PFs) \cite{Iskakov2021}.  These modal energy definitions are applied in power system studies by relating modal quantities to physical energy.  For example, eigenvector-based modal energy is used to evaluate system stability \cite{Hamdan1986}, PF-based modal energy is used to represent modal damping \cite{Chen2014, Hu2022}, and the sum of modal energies is used to define a total action for multi-mode oscillation control \cite{Silva2018}.

All definitions above come with limitations and may show inconsistencies.  
% The application of available modal energy definitions remains unclear.  
For example, under the PF-based definition, the sum of modal energies does not strictly represent the total system damping energy \cite{Chatterjee2022}, which may lead to improper control design when modal energy is used as a control objective.  Another example is reported in \cite{Silva2018}, where modal energy can be complex values, which is inconsistent with physical energy that must be real valued.

% To address the potential confusion arising from the various definitions of modal energy, 

This letter clarifies the applicability of different modal energy definitions to power system dynamics by examining: (i) the condition under which modal energy can be mapped to eigenvalues; (ii) the condition under which modal energy is guaranteed to be real valued; and (iii) the condition under which modal energy can be related to the total system energy.  It is shown that these conditions hold only under restrictive cases, such as when the system state matrix is normal.  

% , which limits the applicability of modal energy approaches in future inverter-dominated power systems.

\vspace{-2mm}
\section{Definitions of Modal Energy}
\label{sec:Def}

The starting point is a the usual nonlinear formulation as an ODE:
\begin{equation}
 \label{eq:DAE}
    \dot{\bs{\xi}} = \bs{f}(\bs{\xi}) .
\end{equation}
An equilibrium point of \eqref{eq:DAE}, say $\bs{\xi}_o$, satisfies $\bs{0} = \bs{f}(\bs{\xi}_o)$.
%
%\begin{equation}
% \label{eq:DAEeq}
%    \bs{0} = \bs{f}(\bs{\xi}_o) .
%\end{equation}
%
Linearization at $\bs{\xi}_o$ leads to:
\begin{equation}\label{eq:LinSys}
    \dot{\bs{x}} = \bs{A} \bs{x} ,
\end{equation}
where $\bs{x} = \bs{\xi} - \bs{\xi}_o$ and $\bs{A} = \nabla^T \bs{f}|_{\bs{\xi}_o}$ is the state matrix of the system.  Assume that $\bs{A}$ has eigenvalues $\bar{\lambda}_i$ with corresponding right eigenvectors $\bar{\bs{v}}_i$ and left eigenvectors $\bar{\bs{u}}_i$, where the overbar $\bar{(\cdot)}$ indicates that these quantities may be complex.  They satisfy
\begin{subequations}
\begin{equation}\label{eq:LefEig}
    \bs{A}\bar{\bs{v}}_i = \bar{\lambda}_i \bar{\bs{v}}_i,
\end{equation}
\begin{equation}\label{eq:RigEig}
    \bar{\bs{u}}_i^{T}\bs{A} = \bar{\lambda}_i \bar{\bs{u}}_i^{T}.
\end{equation}
\end{subequations}
Define matrices $\bar{\bs{V}} = [\bar{\bs{v}}_1 \ \cdots \ \bar{\bs{v}}_n]$ and $\bar{\bs{U}} = [\bar{\bs{u}}_1 \ \cdots \ \bar{\bs{u}}_n]$.  The following identity, used in the next section, holds
\begin{equation}\label{eq:SumFac}
    \bs{V} \bs{U}^{T} = \bs{I}.
    % \sum_{\forall i} \bar{\bs{v}}_{i} \bar{\bs{u}}_{i}^T = \bs{I}.
\end{equation}

The total energy and power of \eqref{eq:LinSys} are expressed as
\begin{subequations}\label{eq:ModDef}
\begin{equation}
    V \! \left(\bs{x}\right) = \frac{1}{2} \bs{x}^T \bs{P} \bs{x},
\end{equation}
\begin{equation}
    \dot{V} \! \left(\bs{x}\right) = \bs{x}^T \bs{P} \dot{\bs{x}},
\end{equation}
\end{subequations}
where $\bs{P}$ is a positive definite matrix.  Alternatively, a normalized energy is obtained by choosing $\bs{P}=\bs{I}$ in \eqref{eq:ModDef}.  
% In power system applications, $V(\bs{x})$ represents the total system energy, while $\dot{V}(\bs{x})$ represents the energy dissipation or energy flow.  

\begin{table*}[!t]
    \caption{Modal energy definitions and properties}
    \vspace{-2mm}
    \centering
    \renewcommand{\arraystretch}{1.5} % Vertical
    \setlength{\tabcolsep}{1.9pt} % Horizontal
    \begin{tabular}{cccp{2cm}p{1.2cm}cp{1.2cm}p{2.2cm}cp{3.0cm}p{2.0cm}cp{2.2cm}p{2.2cm}}
    
        \hline
        \mcii{Method} & & \mcii{Moving Frame-Based} & & \mcii{Eigenvector-Based} & & \mcii{Hermitian PF-Based} & & \mcii{Transpose PF-Based} \\
        \cline{1-2}\cline{4-5}\cline{7-8}\cline{10-11}\cline{13-14}
        
        \multirow{3}{*}{\rotatebox{90}{Definitions}} & \mci{Energy Type} & & \mci{Normalized} & \mci{Physical} &  & \mci{Normalized} & \mci{Physical} & & \mci{Normalized} & \mci{Physical} & & \mci{Normalized} & \mci{Physical} \\
        % \cline{2-2}\cline{4-5}\cline{7-7}\cline{9-10}\cline{12-13}
        
         & Modal Energy & 
        & \mci{$e = \frac{1}{2} \bs{x}^T \bs{x}$} 
        & \mci{$e = \frac{1}{2} \bs{x}^T \bs{P} \bs{x}$} & 
        & \mci{$\bar{e}_i = \frac{1}{2} \bs{x}^T \bar{\bs{v}}_{i} \bar{\bs{u}}_{i}^T \bs{x}$}
        & \mci{$\bar{e}_i = \frac{1}{2} \bs{x}^T \bs{P} \bar{\bs{v}}_{i} \bar{\bs{u}}_{i}^T \bs{x}$} &
        & \mci{$e_i = \frac{1}{2} \bar{\bs{z}}_i^H \bar{\bs{z}}_i $}  
        & \mci{$e_i = \frac{1}{2} \bar{\bs{z}}_i^H \bs{P} \bar{\bs{z}}_i $} & 
        & \mci{$\bar{e}_i = \frac{1}{2} \bar{\bs{z}}_i^T \bar{\bs{z}}_i $} 
        & \mci{$\bar{e}_i = \frac{1}{2} \bar{\bs{z}}_i^T \bs{P} \bar{\bs{z}}_i $} \\
        
         & Modal Power & 
        & \mci{$s_i = \bs{\psi}_i^T \dot{\bs{x}}$} 
        & \mci{$s_i = \bs{\psi}_i^T \bs{P} \dot{\bs{x}}$} &
        & \mci{$\bar{s}_i = \bs{x}^T \bar{\bs{v}}_{i} \bar{\bs{u}}_{i}^T \dot{\bs{x}}$} 
        & \mci{$\bar{s}_i = \bs{x}^T \bs{P} \bar{\bs{v}}_{i} \bar{\bs{u}}_{i}^T \dot{\bs{x}}$} & 
        & \mci{$\bar{s}_i = \bar{\bs{z}}_i^H \dot{\bar{\bs{z}}}_i $} 
        & \mci{$\bar{s}_i = \bar{\bs{z}}_i^H \bs{P} \dot{\bar{\bs{z}}}_i $} & 
        & \mci{$\bar{s}_i = \bar{\bs{z}}_i^T \dot{\bar{\bs{z}}}_i $} 
        & \mci{$\bar{s}_i = \bar{\bs{z}}_i^T \bs{P} \dot{\bar{\bs{z}}}_i $} \\
        \cline{1-2}\cline{4-5}\cline{7-8}\cline{10-11}\cline{13-14}
        
        \multirow{3}{*}{\rotatebox{90}{Properties}} & Eigenvalue & 
        & \mcii{$s_i \neq 2 \bar{\lambda}_i e$} & 
        & \mcii{$\bar{s}_i = 2 \bar{\lambda}_i \bar{e}_i$ always} & 
        & \mcii{$\bar{s}_i = 2 \bar{\lambda}_i e_i$ always} & 
        & \mcii{$\bar{s}_i = 2 \bar{\lambda}_i \bar{e}_i$ always} \\
        
         & Energy Value & 
        & \mcii{$e$ is real} & 
        & \mcii{$\bar{e}_i$ is complex} & 
        & \mcii{$e_i$ is real} & 
        & \mcii{$\bar{e}_i$ is complex} \\
         
         & Energy Sum & 
        & \mcii{$e \equiv V \! \left(\bs{x}\right)$} &
        & \mcii{$\sum_{\forall i} \bar{e}_i = V \! \left(\bs{x}\right)$} & 
        & \mcii{$\sum_{\forall i} e_i \neq V \! \left(\bs{x}\right)$} & 
        & \mcii{$\sum_{\forall i} e_i \neq V \! \left(\bs{x}\right)$} \\
        \hline
        % \multicolumn{13}{p{14.1cm}}{* When the physical energy definition is used, modal energy is a real and positive quantity.} \\
        
    \end{tabular}
    \label{tab:DefTab}
    \vspace{-5mm}
\end{table*}

To connect energy in \eqref{eq:ModDef} with modal analysis, several definitions of modal energy are introduced \cite{MacFarlane1969, Hamdan1986, Iskakov2021}.  The mainstream definitions, denoted by $e_i$ and $s_i$ for modal energy and power, are summarized in Table~\ref{tab:DefTab} and include four types.  The \textit{moving frame-based definition} uses an orthogonal moving frame $\bs{\psi}_i$.  The \textit{eigenvector-based definition} uses both left and right eigenvectors.  The \textit{Hermitian PF-based} and \textit{transpose PF-based definitions} use a new state vector constructed as
\begin{equation}\label{eq:zexpress}
    \bar{\bs{z}}_{i} = \bar{\bs{v}}_{i} \bar{\bs{u}}_{i}^{T} \bs{x}.
\end{equation}

\vspace{-5mm}
\section{Modal Energy Requirements}

% In power system applications, $V(\bs{x})$ in \eqref{eq:ModDef} represents the total system energy, while $\dot{V}(\bs{x})$ represents the energy dissipation or energy flow.  

% The four definitions in Section \ref{sec:Def} are adopted for specific power system problems, such as small-signal stability analysis \cite{Hamdan1986}, energy flow analysis \cite{Chen2014, Hu2022}, and oscillation control \cite{Silva2018}.  

To be consistent with physical systems, such as electric grids, namely systems that must undergo the principles of thermodynamics, modal energy should satisfy the following three conditions:
\begin{itemize}
    \item \textit{Domain of modal energy}: Modal energy should be a real-valued quantity to preserve the physical meaning of energy in power systems.
    \item \textit{Mapping between modal and total energy}: The sum of all modal energies should equal the system total energy $V(\bs{x})$ defined in \eqref{eq:ModDef}, so that modal energy provides a complete representation of power system energy.
\end{itemize}

Moreover, the following feature is desirable:
\begin{itemize}
\item \textit{Mapping between eigenvalues and modal energies}: As eigenvalues are well-established indices, it is desirable to establish a clear relationship between each eigenvalue and the corresponding modal energy.
\end{itemize}

In the following, the four modal energy definitions introduced in Section \ref{sec:Def} are revisited in view of the three requirements above.  Main features and properties are summarized in Table~\ref{tab:DefTab}.

\subsubsection{Moving-Frame-Based} 

Regarding the relationship between eigenvalue and modal energy, we have
\begin{equation}
    \frac{s_{i}}{e} 
    % = \frac{\bs{x}^T \bs{P} \dot{\bs{x}}}{\frac{1}{2} \bs{x}^T \bs{P} \bs{x}}
    = \frac{\bs{x}^T \bs{P} \bs{A} \bs{x}}{\frac{1}{2} \bs{x}^T \bs{P} \bs{x}}
    \overset{\text{when } \bs{x} = \bar{\bs{v}}}{=} 
    % = (\text{for }\bs{x} = \bs{v}) 
    \frac{\bar{\bs{v}}^T \bs{P} \bar{\lambda}_i \bar{\bs{v}}}{\frac{1}{2} \bar{\bs{v}}^T \bs{P} \bar{\bs{v}}}
    = 2 \bar{\lambda}_i \frac{\bar{\bs{v}}^T \bs{P} \bar{\bs{v}}}{\bar{\bs{v}}^T \bs{P} \bar{\bs{v}}}
    =2\bar{\lambda}_i,
\end{equation}
where \eqref{eq:LefEig} is used.

As shown above, this mapping holds only when the state $\bs{x}$ coincides with an eigenvector $\bar{\bs{v}}_i$.  This means that this definition is useful only under specific operating conditions of power systems, which restricts its applicability.  On the other hand, the resulting modal energy is real valued and coincides with the system actual energy defined in \eqref{eq:ModDef}, which is a desirable property.

\subsubsection{Eigenvector-Based} The mapping between eigenvalue and modal energy always holds since
\begin{equation}\label{eq:LefMap}
    \frac{\bar{s}_i}{e_{i}} 
    % \! = \! \frac{\bs{x}^T \bar{\bs{v}}_{i} \bar{\bs{u}}_{i}^T \dot{\bs{x}}}{\frac{1}{2} \bs{x}^T \bar{\bs{v}}_{i} \bar{\bs{u}}_{i}^T \bs{x}}
    \! = \! \frac{\bs{x}^T \bar{\bs{v}}_{i} \bar{\bs{u}}_{i}^T \bs{A} \bs{x}}{\frac{1}{2} \bs{x}^T \bar{\bs{v}}_{i} \bar{\bs{u}}_{i}^T \bs{x}} 
    \! = \! \frac{\bs{x}^T \bar{\bs{v}}_{i} \bar{\lambda}_i \bar{\bs{u}}_{i}^T \bs{x}}{\frac{1}{2} \bs{x}^T \bar{\bs{v}}_{i} \bar{\bs{u}}_{i}^T \bs{x}} 
    \! = \! 2 \bar{\lambda}_i \frac{\bs{x}^T \bar{\bs{v}}_{i} \bar{\bs{u}}_{i}^T \bs{x}}{\bs{x}^T \bar{\bs{v}}_{i} \bar{\bs{u}}_{i}^T \bs{x}} 
    \! = \! 2 \bar{\lambda}_i,\! \!
\end{equation}
where \eqref{eq:RigEig} is used.

The sum of all modal energies equals the total energy
\begin{equation}
    \sum_{\forall i} \bar{e}_i 
    \! = \! \sum_{\forall i} \frac{1}{2} \bs{x}^T \bar{\bs{v}}_{i} \bar{\bs{u}}_{i}^T \bs{x}
    \! = \! \frac{1}{2} \bs{x}^T \left(\sum_{\forall i} \bar{\bs{v}}_{i} \bar{\bs{u}}_{i}^T\right) \bs{x}
    \! = \! V \! \left(\bs{x}\right)\! ,\! \! \! 
\end{equation}
where \eqref{eq:SumFac} is used.
Although the above two conditions are satisfied, the modal energy $\bar{e}_i$ is generally complex-valued, which makes its physical meaning unclear in power systems.

\subsubsection{Hermitian PF-Based} By a derivation similar to \eqref{eq:LefMap}, the eigenvalue-modal energy mapping also holds.  Due to the Hermitian operator, the modal energy is real valued.  However, the sum of modal energies does not generally equal the total energy, and this equality holds only when
\begin{equation}
\begin{aligned}
    \sum_{\forall i} e_i = & \sum_{\forall i} \frac{1}{2} \bar{\bs{z}}_i^H \bs{P} \bar{\bs{z}}_i = \sum_{\forall i} \frac{1}{2} \bs{x}^T \bar{\bs{u}}_{i}^{*} \bar{\bs{v}}_{i}^H \bs{P} \bar{\bs{v}}_{i} \bar{\bs{u}}_{i}^{T} \bs{x} \\
    \overset{\bs{A}^{\sharp} \text{ is normal}}{=} & \frac{1}{2} \bs{x}^T \sum_{\forall i} \left|\bar{\bs{v}}_{i}\right|^2 \left(\bar{\bs{u}}_{i}^{*} \bs{P} \bar{\bs{u}}_{i}^{T}\right) \bs{x} = V \! \left(\bs{x}\right).
\end{aligned}
\end{equation}
This condition requires $\bs{A}$ to satisfy a normality property.  When physical energy is considered, this corresponds to $\bs{A}\bs{A}^{\sharp}=\bs{A}^{\sharp}\bs{A}$, where $\bs{A}^{\sharp}:=\bs{P}^{-1}\bs{A}^{T}\bs{P}$.  For normalized energy, the condition reduces to $\bs{A}\bs{A}^{T}=\bs{A}^{T}\bs{A}$.

The normality of $\bs{A}$ severely restricts its applicability in power systems.  It is possible to show that for second-order synchronous generators (SGs), $\bs{A}^{\sharp}$ is normal only if the network is lossless and machines have no damping.  Moreover, matrix $\bs{A}$ is normal only if all generators are dynamically decoupled, a condition that clearly does not represent an interconnected multi-generator system.  

% Although the Appendix illustrates a simple case, one can infer that this condition becomes increasingly difficult to satisfy in more complex systems.

This mismatch arises from a missing component, namely the cross energy between different modes.  Since $\bs{x}=\sum_{\forall i}\bar{\bs{z}}_i$ from \eqref{eq:zexpress}, the total energy can be expressed as
\begin{equation}
    V \! \left(\bs{x}\right) 
    \! = \! \sum_{\forall i,j} \frac{1}{2} \bar{\bs{z}}_i^H \bs{P} \bar{\bs{z}}_j 
    \! = \! \underbrace{\sum_{\forall i} \frac{1}{2}\,\bar{\bs{z}}_i^{H}\bs{P}\bar{\bs{z}}_i}_{\text{modal energy}} \! + \! \underbrace{\sum_{i \neq j} \frac{1}{2}\,\bar{\bs{z}}_{i}^{H}\bs{P}\bar{\bs{z}}_{j}}_{\text{missing energy}}.\! 
\end{equation}

\subsubsection{Conjugate PF-Based} The conclusions are similar to the Hermitian PF-based method, except that the modal energy is complex valued rather than real.

\vspace{-3mm}
\section{Case Study}

A WECC 9-bus system is used for demonstration.  Two scenarios, one with only SGs and the other including inverters, are considered.  For both scenarios, we consider the linearized system at a given equilibrium point and then apply a step variations to the variables in order to emulate the effect of a short circuit near Bus 7.  Then, we assume that the fault is cleared without changing the topology of the system.  In this way, we can calculate the trajectory of the state vector $\bs{x}$ and compare the dynamic behavior of the modal energies obtained with the four definitions discussed in the previous sections. 

% is obtained from the fully-fledged nonlinear time-domain simulation, while the eigenvalues and eigenvectors are obtained from the pre-disturbance linearized model, which changes little after the disturbance, as both  network topology and final operating point are same as the pre-disturbance one.

Considering, for sake of example, $t=1.5$ s after applying the disturbance, for the case with only 2nd-order SGs, the total energy defined in \eqref{eq:ModDef} is $1.43$, and the corresponding total dissipated power is $-4.86$.  The modal energy results reported in Table~\ref{tab:Bus9Mod} are consistent with theory.  The moving-frame-based method preserves the system actual energy, and its first term of modal power represents the actual dissipated power; however, the mapping between eigenvalue and modal energy does not exist.  The eigenvector-based method satisfies the eigenvalue mapping and energy additivity, but the modal energy can be complex valued.  The Hermitian PF-based method preserves eigenvalue mapping and has real-valued energy, but the summed modal energy is $1.02$, which is smaller than the total energy due to the non-normality of the system.  The transpose PF-based method produces complex-valued energy, and its sum is $0.31$, which deviates significantly from the system actual energy.

\begin{table*}[t]
    \caption{Modal Energy of WECC 9-Bus System}
    \vspace{-2mm}
    \centering
    \renewcommand{\arraystretch}{1.2} % Vertical
    \setlength{\tabcolsep}{1.0pt} % Horizontal
    \begin{tabular}{ccccccccccccc}
    
        \hline
        \multirow{2}{*}{Eigenvalues} & & \mcii{Moving Frame-Based} & & \mcii{Eigenvector-Based} & & \mcii{Hermitian PF-Based} & &  \mcii{Transpose PF-Based} \\
        
        \cline{3-4}\cline{6-7}\cline{9-10}\cline{12-13}
        &  & Energy & Power  & & Energy & Power  & & Energy & Power  & & Energy & Power  \\
         
        \cline{1-1}\cline{3-4}\cline{6-7}\cline{9-10}\cline{12-13}
        $-0.73\pm \uj12.17$ & & \multirow{4}{*}{$1.43$} & 
        \multirow{4}{*}{%
        \makecell[{{p{2cm}}}]{\raggedright
        Active: $-4.86$;\\
        Reactive: $80.56$, $27.45$, $-8.87$, $41.24$, $20.68$}} & & 
        $0.005 \mp \uj 0.001$ & $0.02\pm \uj 0.12$ & &
        $2\times0.006$ & $-0.01\pm \uj0.13$ & & $0.0002\mp \uj 0.0003$ & $0.006 \pm \uj 0.005$ \\
        
        $-0.38\pm\uj7.91$ & & & & & 
        $0.55\pm\uj0.12$ & $-2.30\pm\uj8.56$ & &
        $2\times0.37$ & $-0.29\pm\uj5.78$ & & $0.02\mp\uj0.01$ & $0.07\pm\uj0.26$ \\
        
        $-0.46$ & & & & & 
        $0.33$ & $-0.30$ & &
        $0.28$ & $-0.26$ & & $0.28$ & $-0.26$ \\
        
        $0$ & & & & & 
        $0$ & $0$ & &
        $0$ & $0$ & & $0$ & $0$ \\
        \hline
        % \cline{1-1}\cline{3-4}\cline{6-7}\cline{9-10}
        
         & & $e=V \! \left(\bs{x}\right)$ 
         & $s_i \neq 2 \bar{\lambda}_i e$ & 
         & $\sum_{\forall i} \bar{e}_i = V \! \left(\bs{x}\right)$ 
         & $\bar{s}_i = 2 \bar{\lambda}_i \bar{e}_i$ & 
         & $\sum_{\forall i} e_i \neq V \! \left(\bs{x}\right)$ 
         & $\bar{s}_i = 2 \bar{\lambda}_i e_i$ & 
         & $\sum_{\forall i} \bar{e}_i \neq V \! \left(\bs{x}\right)$
         & $\bar{s}_i = 2 \bar{\lambda}_i \bar{e}_i$ \\
        \cline{2-13}
                
    \end{tabular}
    \vspace{-0.4cm}
    \label{tab:Bus9Mod}
\end{table*}

\begin{figure}[!t]
    \centering
    \subfloat{%
        \includegraphics[width=2.8in]{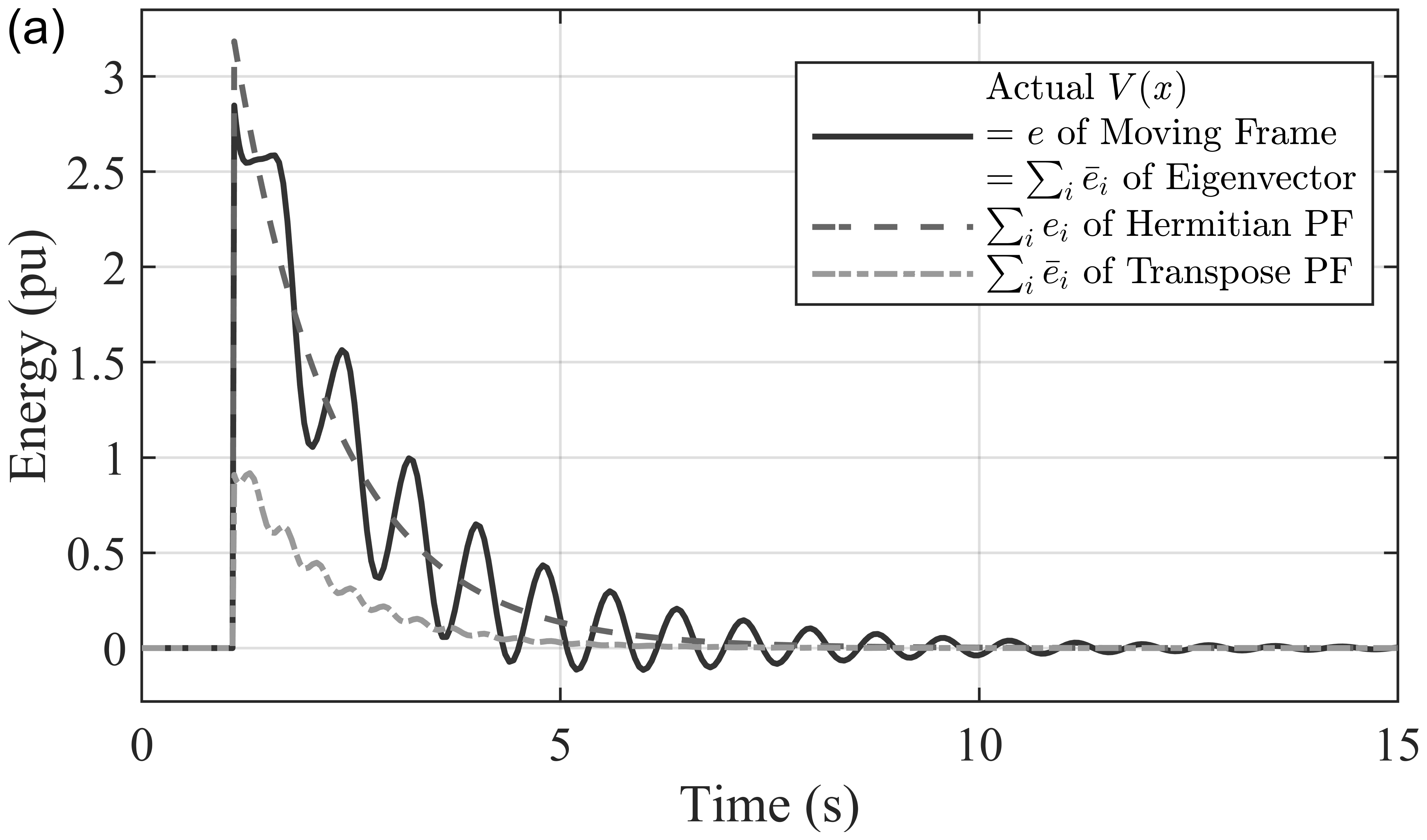}%
        \label{fig:Bus9Ene}}
    \\[4pt]
    \subfloat{%
        \includegraphics[width=2.8in]{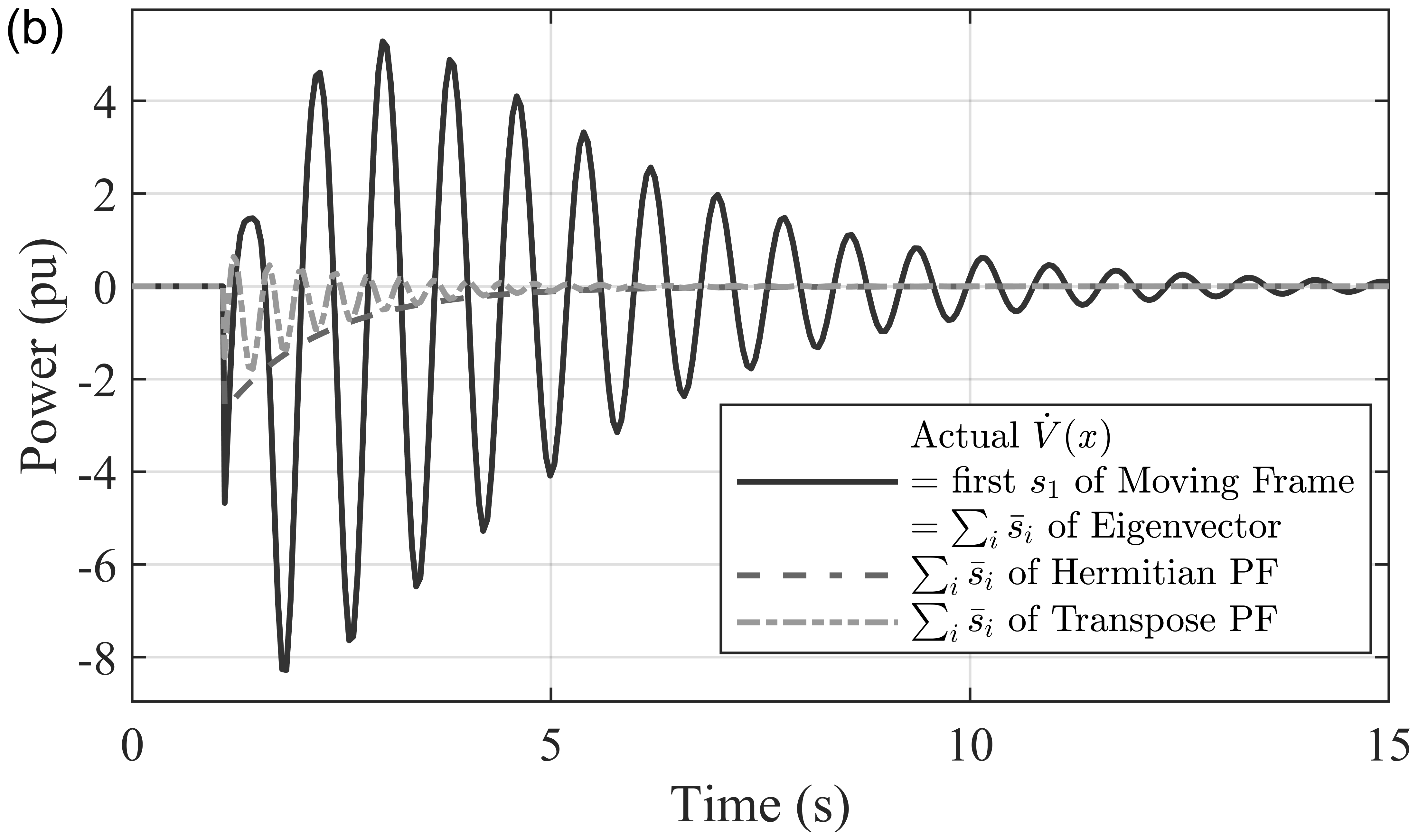}%
        \label{fig:Bus9Pow}}
\vspace{-2mm}
        \caption{Dynamics of (a) stored energy and (b) dissipated power.}
    \label{fig:Bus9}
\vspace{-3mm}
\end{figure}

Energy and power trajectories are shown in Fig.~\ref{fig:Bus9} assuming that the disturbance on the states is applied at $t = 1$ s.  As expected, for the moving-frame-based method, the modal energy equals the actual energy.  For the eigenvector-based method, both the energy sum and power sum match the total energy and power.  For the Hermitian PF-based and transpose PF-based methods, the summed modal energy is lower than the total energy, and the summed modal power deviates significantly from the actual value.  In particular, even the sign of the power, which indicates stability (negative) or instability (positive) as discussed in \cite{Chen2014}, may be incorrect.  This leads to an incorrect assessment of power system stability.

The Hermitian PF-based method is further examined by including exciters and DFIG wind generators.  For visualization, system normality is quantified using the following index, where a value of $1$ indicates perfect normality:
\begin{equation}
    \mathrm{Normality}(\bs{A}) =
    \frac{1}{1 + \lVert \bs{A}^{\sharp} \bs{A} - \bs{A} \bs{A}^{\sharp} \rVert_{F} / \lVert \bs{A} \rVert_{F}^{2}},
\end{equation}
where $\lVert \cdot \rVert_{F}$ denotes the Frobenius norm.

\begin{table}[!t]
    \centering
    \setlength{\tabcolsep}{2.2pt} % Horizontal
    \caption{Hermitian PF-Based Method Regarding Summed Energy}
    \vspace{-2mm}
    \renewcommand{\arraystretch}{1.2} % Vertical
    \label{tab:MorSys}
    \begin{tabular}{c c r r r r}
        \hline
        \mcii{System} &
        \makecell[r]{No-damping SG and \\ Lossless Network} &
        \makecell[r]{2nd-order\\SG} &
        \makecell[r]{SG with\\Exciter$^*$} &
        DFIG$^*$ \\
        \hline
        \mcii{Normality($\bs{A}$)} & $1$ & $0.74$ & $0.63$ & $0.58$ \\
        \hline
        \multirow{3}{*}{Energy} & Actual & $1.93$ & $1.43$   & $233$ & $304$  \\
                                % \cline{2-5}
                                & Summed Modal & $1.93$ & $1.02$   & $1094$ & $2872$ \\
                                % \cline{2-5}
                                & Error (\%) & $0$ & $29$ & $369$ & $845$ \\
        \hline
        \multicolumn{6}{p{8.1cm}}{* Normalized energy is used since it is difficult to define $\bs{P}$ here.} \\
    \end{tabular}
    \vspace{-7mm}
\end{table}

Table~\ref{tab:MorSys} shows the system normality and energy calculated again at $t=1.5$ s after the contingency. For comparison, the system with no damping and a lossless network has perfect normality as proved in Section III, and the error in the summed modal energy is zero. With regular SGs, a normality value of $0.74$ leads to a $29\%$ error in energy. If the exciter or the DFIG wind generator is included, system normality decreases due to increased model complexity, and the energy error increases significantly.  Thus Hermitian PF-based method has limited applicability in realistic models of power systems.

\vspace{-3mm}
\section{Conclusions}

This letter clarifies the definitions and applicability of mainstream modal energy approaches in power system analysis, including the moving frame–based, eigenvector–based, Hermitian PF–based, and transpose PF–based methods, and reveals their limitations in complex and inverter-dominated power systems.  The moving frame–based method lacks a mapping between modal energy and eigenvalue.  The eigenvector–based method yields complex-valued modal energy, which contradicts physical energy definitions.  PF-based methods fail to preserve energy additivity, which holds only when the state matrix is normal.

\vspace{-2mm}
\bibliography{refs}

@article{MacFarlane1969,
  author  = {A. G. J. MacFarlane},
  title   = {Use of power and energy concepts in the analysis of multivariable feedback controllers},
  journal = {Proc. Inst. Electr. Eng.},
  volume  = {116},
  number  = {8},
  pages   = {1449--1452},
  year    = {1969},
  doi     = {10.1049/piee.1969.0261}
}

@article{Iskakov2021,
  author  = {A. B. Iskakov},
  title   = {Definition of State-in-Mode Participation Factors for Modal Analysis of Linear Systems},
  journal = {IEEE Trans. Autom. Control},
  volume  = {66},
  number  = {11},
  pages   = {5385--5392},
  year    = {2021},
  doi     = {10.1109/TAC.2020.3043312}
}

@inproceedings{Silva2018,
  author    = {H. Silva-Saravia and Y. Wang and H. Pulgar-Painemal and K. Tomsovic},
  title     = {Oscillation energy based sensitivity analysis and control for multi-mode oscillation systems},
  booktitle = {PES General Meeting},
  pages     = {1--5},
  year      = {2018},
  doi       = {10.1109/PESGM.2018.8586117}
}

@article{Chen2014,
  author  = {L. Chen and Y. Min and Y.-P. Chen and W. Hu},
  title   = {Evaluation of Generator Damping Using Oscillation Energy Dissipation and the Connection With Modal Analysis},
  journal = {IEEE Trans. Power Syst.},
  volume  = {29},
  number  = {3},
  pages   = {1393--1402},
  year    = {2014},
  doi     = {10.1109/TPWRS.2013.2291898}
}

@article{Hamdan1986,
  author  = {A. M. A. Hamdan},
  title   = {Coupling measures between modes and state variables in power-system dynamics},
  journal = {Int. J. Control},
  volume  = {43},
  number  = {3},
  pages   = {1029--1041},
  year    = {1986},
  doi     = {10.1080/00207178608933521}
}

@article{Chatterjee2022,
  author  = {K. Chatterjee and N. R. Chaudhuri},
  title   = {On the Equality of Modal Damping Power and the Average Rate of Transient Energy Dissipation in a Multimachine Power System},
  journal = {IEEE Control Syst. Lett.},
  volume  = {6},
  pages   = {1531--1536},
  year    = {2022},
  doi     = {10.1109/LCSYS.2021.3121008}
}

@book{Gibbard2015,
  author  = {M. J. Gibbard and P. Pourbeik and D. J. Vowles},
  title   = {Small-Signal Stability, Control and Dynamic Performance of Power Systems},
   publisher = {University of Adelaide Press},
  address   = {Adelaide},
  year    = {2015}
}

@book{Milano2021,
  author  = {F. Milano and I. Dassios and M. Liu and G. Tzounas},
  title   = {Eigenvalue Problems in Power Systems},
   publisher = {CRC Press},
  address   = {Boca Raton},
  year    = {2021}
}

@article{Hu2022,
  author  = {Y. Hu and S. Bu and S. Yi and J. Zhu and J. Luo and Y. Wei},
  title   = {A Novel Energy Flow Analysis and Its Connection With Modal Analysis for Investigating Electromechanical Oscillations in Multi-Machine Power Systems},
  journal = {IEEE Trans. Power Syst.},
  volume  = {37},
  number  = {2},
  pages   = {1139--1150},
  year    = {2022},
  doi     = {10.1109/TPWRS.2021.3099474}
}

\end{document}